\documentclass[aps,pra,floatfix,showpacs,10pt,twocolumn]{revtex4}
\usepackage[dvips]{graphicx}
\usepackage{amsmath,amsfonts,amssymb,graphics,graphicx,epsfig,color,times,bbm}

\begin{document}

\title{Thermal entanglement in the anisotropic Heisenberg XXZ model with the Dzyaloshinskii-Moriya interaction}
\author{Da-Chuang Li$^{1}$\footnote{E-mail: dachuang@ahu.edu.cn}, Xian-Ping Wang$^{1}$, Zhuo-Liang Cao$^{1,
2}$\footnote{E-mail:zhuoliangcao@gmail.com (Corresponding
Author)}}

\affiliation{$^{1}$Key Laboratory of Opto-electronic Information
Acquisition and Manipulation (Ministry of Education), School of
Physics {\&} Material Science, Anhui
University, Hefei 230039 P. R. China\\
$^{2}$Department of Physics, Hefei Teachers College, Hefei 230061
P. R. China}



\begin{abstract}
The entanglement is investigated in a two-qubit Heisenberg XXZ
system with Dzyaloshinskii-Moriya(DM) interaction. It is shown
that the entanglement can be efficiently controlled by the DM
interaction parameter and coupling coefficient $J_{z}$.
$D_{x}$(the x-component parameter of the DM interaction) has a
more remarkable influence on the entanglement and the critical
temperature than $D_{z}$(the z-component parameter of the DM
interaction). Thus, by changing the direction of DM interaction,
we can get a more efficient control parameter to increase the
entanglement and the critical temperature.
\end{abstract}

\maketitle

\section{introduction}
Entanglement has been extensively studied in recent years because
it has the fascinating nonclassical nature of quantum mechanics,
and it plays a key role in quantum information processing
\cite{1,2}. The quantum entanglement in solid state systems such
as spin chains is an important emerging field \cite{3,4,5,6}, spin
chains are natural candidates for the realization of entanglement,
and spin has been researched in many other systems, such as
superconductor \cite{9,10}, quantum dots \cite{11,12,13}, and
trapped ion \cite{14,15}.

In order to characterize qualitatively and quantitatively the
entanglement properties of condensed matter systems and apply them
in quantum information, the thermal entanglement qualities in
Heisenberg model have been extensively studied
\cite{7,8,27,28,29}, and many schemes of teleportation via thermal
entangled state have been reported \cite{30,20,21,22,23}. In
condensed matter systems, the Heisenberg chains have also been
used to construct a quantum computer \cite{16}, perform quantum
computation \cite{17,18,19}, and so on.

In those studies the spin-spin interaction was considered, but the
spin-orbit coupling was rarely considered. Especially the
influences of x-component DM interaction parameter on the
entanglement and the critical temperature have never been
reported. In this paper we investigate the influence of
Dzyaloshinskii-Moriya interaction parameter (arising from the
spin-orbit coupling) and coupling coefficient $J_{z}$ on the
entanglement of a two-qubit anisotropic Heisenberg XXZ spin chain.
We show that the DM interaction parameter and the coupling
coefficient $J_{z}$ are both efficient control parameters of
entanglement, increasing them can enhance the entanglement or slow
down the decreace of the entanglement. In addition, by analyzing
we know that different component parameters of DM interaction have
different influences on the entanglement and the critical
temperature $T_{c}$, the parameter $D_{x}$ (x-component parameter
of the DM interaction) has a more remarkable influences than the
parameter $D_{z}$ (z-component parameter of the DM interaction).
So more efficient control parameter can be obtained by changing
the DM interaction direction.

Our paper is organized as follows. In Section II, we introduce the
Hamiltonian of the two-qubit anisotropic Heisenberg XXZ chain with
the z-component parameter of the DM interaction, we calculate the
concurrence of this system, analyze the influence of parameters on
the entanglement in ground state and thermal state. In Section
III, we similarly analyze the model of the two-qubit Heisenberg
XXZ chain with the x-component parameter of the DM interaction.
Then we compare the influences of the two component parameters of
the DM interaction on the entanglement in Section IV. Finally, in
Section V a discussion concludes the paper.

\section{XXZ Heisenberg model with DM interaction parameter $D_{z}$}
The Hamiltonian $H$ for a two-qubit anisotropic Heisenberg XXZ
chain with DM interaction parameter $D_{z}$ is
\begin{equation}
\label{1}
H=J\sigma_{1}^{x}\sigma_{2}^{x}+J\sigma_{1}^{y}\sigma_{2}^{y}+J_{z}\sigma_{1}^{z}\sigma_{2}^{z}
+D_{z}(\sigma_{1}^{x}\sigma_{2}^{y}-\sigma_{1}^{y}\sigma_{2}^{x}),
\end{equation}
where $J$ and $J_{z}$ are the real coupling coefficients, $D_{z}$
is the z-component parameter of DM interaction,
$\sigma^{i}(i=x,y,z)$ are Pauli matrices. The coupling constants
$J>0$ and $J_{z}>0$ correspond to the antiferromagnetic case,
$J<0$ and $J_{z}<0$ correspond to the ferromagnetic case. This
model is reduced to isotropic XX model when $J_{z}=0$ and
isotropic XXX model when $J_{z}=J$. Parameters $J$, $J_{z}$ and
$D_{z}$ are dimensionless.

In the standard basis
$\{|00\rangle,|01\rangle,|10\rangle,|11\rangle\}$, the Hamiltonian
(1) can be expressed as
\begin{equation}
\label{2} H=\left(
\begin{array}{cccc}
  J_{z} & 0 & 0 & 0 \\
  0 & -J_{z} & 2J+2iD_{z} & 0 \\
  0 & 2J-2iD_{z} & -J_{z} & 0 \\
  0 & 0 & 0 & J_{z} \\
\end{array}
\right).
\end{equation}
By a straightforward calculation we can obtain $H$ eigenstates:
\begin{subequations}\label{3}
\begin{equation}
|\Phi_{1}\rangle=|00\rangle,
\end{equation}
\begin{equation}
|\Phi_{2}\rangle=|11\rangle,
\end{equation}
\begin{equation}
|\Phi_{3}\rangle=\frac{1}{\sqrt{2}}(e^{i\theta}|01\rangle+|10\rangle),
\end{equation}
\begin{equation}
|\Phi_{4}\rangle=\frac{1}{\sqrt{2}}(e^{i\theta}|01\rangle-|10\rangle),
\end{equation}
\end{subequations}
with corresponding eigenvalues:
\begin{subequations}\label{4}
\begin{equation}
E_{1}=J_{z},
\end{equation}
\begin{equation}
E_{2}=J_{z},
\end{equation}
\begin{equation}
E_{3}=-J_{z}+2w,
\end{equation}
\begin{equation}
E_{4}=-J_{z}-2w,
\end{equation}
\end{subequations}
where $w=\sqrt{J^{2}+D^{2}_{z}}$,
$\theta=\arctan(\frac{D_{z}}{J})$.

The state of a spin chain system at thermal equilibrium is
$\rho(T)=\frac{\exp(-\beta H)}{Z}$, where $Z=tr[\exp(-\beta H)]$
is the partition function of the system, $H$ is the system
Hamiltonian and $\beta=\frac{1}{K_{B}T}$, with $T$ is temperature
and $K_{B}$ is the Boltzmann costant which we take equal to 1 for
simplicity. Here $\rho(T)$ represents a thermal state, so the
entanglement in the thermal state is called thermal entanglement
\cite{24}. In the above standard basis, the state of this system
at thermal equilibrium can be expressed as:
\begin{equation}
\label{5} \rho(T)=\frac{1}{Z}\left(
\begin{array}{cccc}
  e^{-\beta J_{z}} & 0 & 0 & 0 \\
  0 & u & ve^{i\theta} & 0 \\
  0 & ve^{-i\theta} & u & 0 \\
  0 & 0 & 0 & e^{-\beta J_{z}} \\
\end{array}\right),
\end{equation}
where $u=\frac{1}{2}(1+e^{4\beta w})e^{\beta(J_{z}-2w)}$,
$v=\frac{1}{2}(1-e^{4\beta w})e^{\beta(J_{z}-2w)}$, and
$Z=2e^{-\beta J_{z}}[1+e^{2\beta J_{z}}\cosh(2\beta w)]$.

In what follows, we consider the concurrence to quantify the
amount of entanglement of the above two-qubit system state
$\rho(T)$. The concurrence \cite{25,26} is defined as
$C(\rho(T))=\max[2\max(\lambda_{i})-\Sigma_{i}\lambda_{i},0]$,
where $\lambda_{i}(i=1,2,3,4)$  are the square roots of the
eigenvalues of the matrix $R=\rho S\rho^{*}S$, in which
$S=\sigma_{1}^{y}\bigotimes\sigma_{2}^{y}$, $\rho$ is the density
matrix of the Eq. (5), and the asterisk  denote the complex
conjugate. After some straightforward calculation, we get
\begin{subequations}\label{6}
\begin{equation}
\lambda_{1}=\frac{1}{Z}e^{-\beta J_{z}},
\end{equation}
\begin{equation}
\lambda_{2}=\frac{1}{Z}e^{-\beta J_{z}},
\end{equation}
\begin{equation}
\lambda_{3}=\frac{1}{Z}e^{\beta(J_{z}-2w)},
\end{equation}
\begin{equation}
\lambda_{4}=\frac{1}{Z}e^{\beta(J_{z}+2w)},
\end{equation}
\end{subequations}
thus the corresponding concurrence can be expressed as:
\begin{equation}
\label{7} C(\rho(T))=\max\{\frac{e^{\beta J_{z}}}{Z}(|e^{2\beta
w}-e^{-2\beta J_{z}}|-e^{-2\beta w}-e^{-2\beta J_{z}}),0\}.
\end{equation}
The concurrence is invariant under the substitutions
$J\rightarrow-J$ and $D_{z}\rightarrow-D_{z}$, so we can restrict
$J>0$ and $D_{z}>0$ without loss of generality. The concurrence
ranges from 0 to 1, $C(\rho(T))=0$ and $C(\rho(T))=1$ indicate the
vanishing entanglement and the maximal entanglement respectively.
We can see from Eq. (7) that the entanglement
$C(\rho(T))=\frac{e^{\beta J_{z}}}{Z}(e^{2\beta w}-e^{-2\beta
J_{z}}-e^{-2\beta w}-e^{-2\beta J_{z}})$ if $J_{z}>-w$, and
$C(\rho(T))=0$ if $J_{z}<-w$. Here we analyze the $J_{z}>-w$ case.

When $T=0$, the system is in its ground state. It is easy to find
that the ground-state energy is equal to
\begin{subequations}\label{8}
\begin{equation}
E_{4}=-J_{z}-2w,     if J_{z}>-w,
\end{equation}
\begin{equation}
E_{1}=E_{2}=J_{z},     if J_{z}<-w.
\end{equation}
\end{subequations}
Thus, the ground state is the disentangled state
$|\Phi_{1}\rangle$ or $|\Phi_{2}\rangle$ when $J_{z}<-w$, and the
ground state is the entangled state $|\Phi_{4}\rangle$ when
$J_{z}>-w$. The entanglement of the ground state
$|\Phi_{4}\rangle$ is the maximal entanglement with
$C(|\Phi_{4}\rangle)=1$.

As the temperature increases, thermal fluctuation will be
introduced into the system, thus the entanglement will be changed
due to the mix of the ground states and the excited states. When
the temperature is higher than a critical temperature the
entanglement is zero. Quantum phase transition happens at the
critical temperature $T_{c}$. From Eq. (7), we obtain the
following critical temperature equation
\begin{equation}\label{9}
e^{\frac{2J_{z}}{T_{c}}}\sinh(\frac{2w}{T_{c}})=1.
\end{equation}
\begin{figure}[tbp]
\includegraphics[scale=0.5,angle=0]{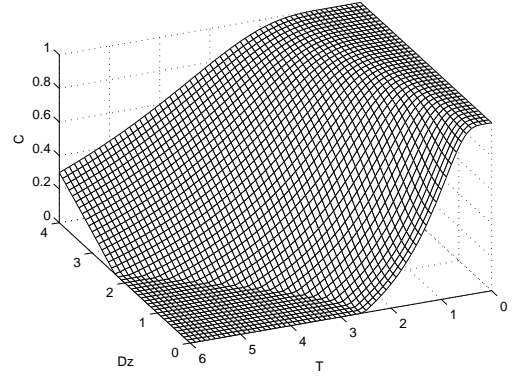}
\caption{The concurrence is plotted versus $T$ and $D_{z}$ where
the coupling constants $J=1$ and $J_{z}=0.2$.} \label{fig1}
\end{figure}

\begin{figure}[tbp]
\includegraphics[scale=0.5,angle=0]{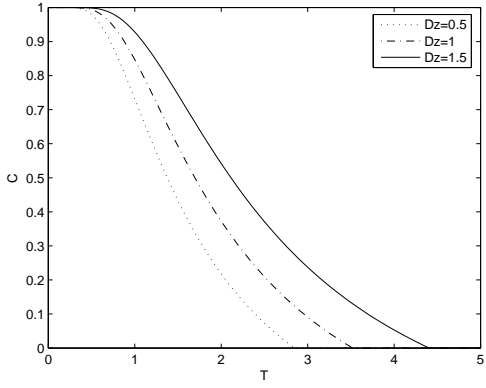}
\caption{The concurrence is plotted as a function of the
temperature $T$ for different $D_{z}$, here $J=1$ and
$J_{z}=0.2$.} \label{fig2}
\end{figure}

\begin{figure}[tbp]
\includegraphics[scale=0.5,angle=0]{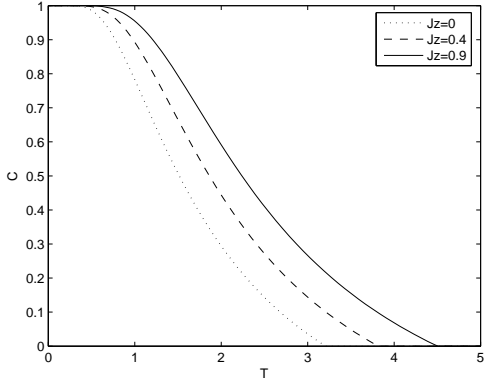}
\caption{The concurrence versus $T$ for different $J_{z}$ in the
system with z-component parameter $D_{z}$, here $J=1$ and
$D_{z}=1$.} \label{fig3}
\end{figure}

To see the change of the entanglement in detail, we analyze the
concurrence of Eq. (7). By fixing some parameters we can know the
roles of the other parameters and the variation of the
entanglement. In Fig. 1, the thermal entanglement is plotted
versus $T$ and $D_{z}$ where the coupling constants $J=1$ and
$J_{z}=0.2$. From the figure, it's obvious that the increased
temperature $T$ can decrease the entanglement. The reason is the
mixing of the maximally entangled state with other states. In
addition, it is easy to see that the entanglement will increase as
the DM interaction parameter $D_{z}$ increases. Fig. 2
demonstrates the concurrence versus temperature for different DM
coupling parameter $D_{z}$ when $J=1$ and $J_{z}=0.2$. It shows
that the concurrence will decrease with increasing temperature $T$
and increase with increasing $D_{z}$ for a certain temperature.
The critical temperature $T_{c}$ determined by Eq. (9) is
dependent on $D_{z}$, increasing $D_{z}$ can increase the critical
temperature above which the entanglement vanishes. Similarly, Fig.
3 shows the concurrence versus temperature for different
anisotropic coupling parameter $J_{z}$ when $J=1$ and $D_{z}=1$.
We can see that the entanglement decrease with the increase of
temperature, and by increasing $J_{z}$, the critical temperature
is increased and the entanglement is enhanced for a certain
temperature.

So the DM interaction parameter $D_{z}$ and anisotropic coupling
coefficient $J_{z}$ are both efficient control parameters of
entanglement, by increasing them, we can enhance the entanglement
or increase the critical temperature to slow down the decreace of
the entanglement.

\section{XXZ Heisenberg model with DM interaction parameter $D_{x}$}
Here we consider the case of the two-qubit anisotropic Heisenberg
XXZ chain with DM interaction parameter $D_{x}$. The Hamiltonian
is
\begin{equation}
\label{10}
H^{'}=J\sigma_{1}^{x}\sigma_{2}^{x}+J\sigma_{1}^{y}\sigma_{2}^{y}+J_{z}\sigma_{1}^{z}\sigma_{2}^{z}
+D_{x}(\sigma_{1}^{y}\sigma_{2}^{z}-\sigma_{1}^{z}\sigma_{2}^{y}),
\end{equation}
where $D_{x}$ is the x-component parameter of DM interaction, $J$,
$J_{z}$ and $\sigma^{i}(i=x,y,z)$ are the same as those in Section
II. Parameters $D_{x}$, $J$ and $J_{z}$ are dimensionless.

In the standard basis
$\{|00\rangle,|01\rangle,|10\rangle,|11\rangle\}$, the Hamiltonian
(10) can be rewritten as
\begin{equation}
\label{11} H^{'}=\left(
\begin{array}{cccc}
  J_{z} & iD_{x} & -iD_{x} & 0 \\
  -iD_{x} & -J_{z} & 2J & iD_{x} \\
  iD_{x} & 2J & -J_{z} & -iD_{x} \\
  0 & -iD_{x} & iD_{x} & J_{z} \\
\end{array}
\right).
\end{equation}
After a straightforward calculation we obtain $H^{'}$ eigenstates:
\begin{subequations}\label{12}
\begin{equation}
|\Psi_{1}\rangle=\frac{1}{\sqrt{2}}(|00\rangle+|11\rangle),
\end{equation}
\begin{equation}
|\Psi_{2}\rangle=\frac{1}{\sqrt{2}}(|01\rangle+|10\rangle),
\end{equation}
\begin{equation}
|\Psi_{3}\rangle=\frac{1}{\sqrt{2}}(-i\sin\phi|00\rangle+\cos\phi|01\rangle-\cos\phi|10\rangle+i\sin\phi|11\rangle),
\end{equation}
\begin{equation}
|\Psi_{4}\rangle=\frac{1}{\sqrt{2}}(-i\sin\varphi|00\rangle+\cos\varphi|01\rangle-\cos\varphi|10\rangle+i\sin\varphi|11\rangle),
\end{equation}
\end{subequations}
with corresponding eigenvalues:
\begin{subequations}\label{13}
\begin{equation}
E_{1}^{'}=J_{z},
\end{equation}
\begin{equation}
E_{2}^{'}=2J-J_{z},
\end{equation}
\begin{equation}
E_{3}^{'}=-J+w^{'},
\end{equation}
\begin{equation}
E_{4}^{'}=-J-w^{'},
\end{equation}
\end{subequations}
where $\phi=\arctan(\frac{2D_{x}}{J+J_{z}-w^{'}})$,
$\varphi=\arctan(\frac{2D_{x}}{J+J_{z}+w^{'}})$, and
$w^{'}=\sqrt{(J+J_{z})^{2}+4D^{2}_{x}}$.

For convenience of analysis, we assume $J_{z}\leqslant J$ in this
section (for the case $J_{z}>J$ we can get some similar results).
Here, it is easy to see that the system ground-state energy is
$E_{4}^{'}=-J-w^{'}$, thus the corresponding ground state
$|\Psi_{4}\rangle$ is the maximally entangled state with
$C(|\Psi_{4}\rangle)=1$. In fact the four eigenstates in Eq. (12)
are all the maximally entangled states, this phenomenon indicates
that the ground states have more entanglement in this system than
the system of Eq. (1).

At thermal equilibrium the density matrix of this two-qubit spin
chain system has the following form:
\begin{eqnarray}\label{14}
\rho^{'}(T)=\frac{exp(-\beta H^{'})}{Z^{'}}=\frac{1}{2Z^{'}}\left(
\begin{array}{cccc}
  \mu_{+} & -\xi & \xi & \mu_{-} \\
  \xi & \nu_{+} & \nu_{-} & -\xi \\
  -\xi & \nu_{-} & \nu_{+} & \xi \\
  \mu_{-} & \xi & -\xi & \mu_{+} \\
\end{array}
\right),
\end{eqnarray}
where $Z^{'}=2e^{-\beta J}\cosh[\beta(J-J_{z})]+2e^{\beta
J}\cosh(\beta w^{'})$ is the partition function of the system,
$H^{'}$ is the system Hamiltonian and $\beta=\frac{1}{K_{B}T}$
with the Boltzmann costant $K_{B}\equiv1$. $\mu_{\pm}=e^{-\beta
J_{z}}\pm(e^{\beta(J-w^{'})}\sin^{2}\phi+e^{\beta(J+w^{'})}\sin^{2}\varphi)$,
$\nu_{\pm}=e^{\beta(J_{z}-2J)}\pm(e^{\beta(J-w^{'})}\cos^{2}\phi+e^{\beta(J+w^{'})}\cos^{2}\varphi)$,
and
$\xi=ie^{\beta(J-w^{'})}\sin\phi\cos\phi+ie^{\beta(J+w^{'})}\sin\varphi\cos\varphi$.

In what follows, we calculate the square roots of the eigenvalues
of the matrix $R^{'}=\rho^{'}S\rho^{'*}S$, where $\rho^{'*}$ is
the complex conjugate of $\rho^{'}$ and
$S=\sigma_{1}^{y}\bigotimes\sigma_{2}^{y}$. The square roots of
the eigenvalues of the matrix $R^{'}$ are:
\begin{subequations}\label{15}
\begin{equation}
\lambda_{1}^{'}=\frac{1}{Z^{'}}e^{\beta(J_{z}-2J)},
\end{equation}
\begin{equation}
\lambda_{2}^{'}=\frac{1}{Z^{'}}e^{-\beta J_{z}},
\end{equation}
\begin{equation}
\lambda_{3}^{'}=\frac{1}{Z^{'}}e^{\beta J}[\cosh(\beta
w^{'})+\sqrt{\cosh^{2}(\beta w^{'})-1}],
\end{equation}
\begin{equation}
\lambda_{4}^{'}=\frac{1}{Z^{'}}e^{\beta J}[\cosh(\beta
w^{'})-\sqrt{\cosh^{2}(\beta w^{'})-1}].
\end{equation}
\end{subequations}
According to the methods in \cite{25,26}, when $J_{z}\leqslant J$
we obtain the corresponding concurrence
\begin{eqnarray}\label{16}
C(\rho^{'}(T))&=&\max\{\frac{1}{Z^{'}}[|e^{\beta J}(\cosh(\beta
w^{'})+\sqrt{\cosh^{2}(\beta w^{'})-1})\nonumber\\
&-&e^{-\beta J_{z}}|-e^{\beta J}(\cosh(\beta
w^{'})-\sqrt{\cosh^{2}(\beta
w^{'})-1})\nonumber\\
&-&e^{\beta(J_{z}-2J)}],0\}.
\end{eqnarray}

\begin{figure}[tbp]
\includegraphics[scale=0.5,angle=0]{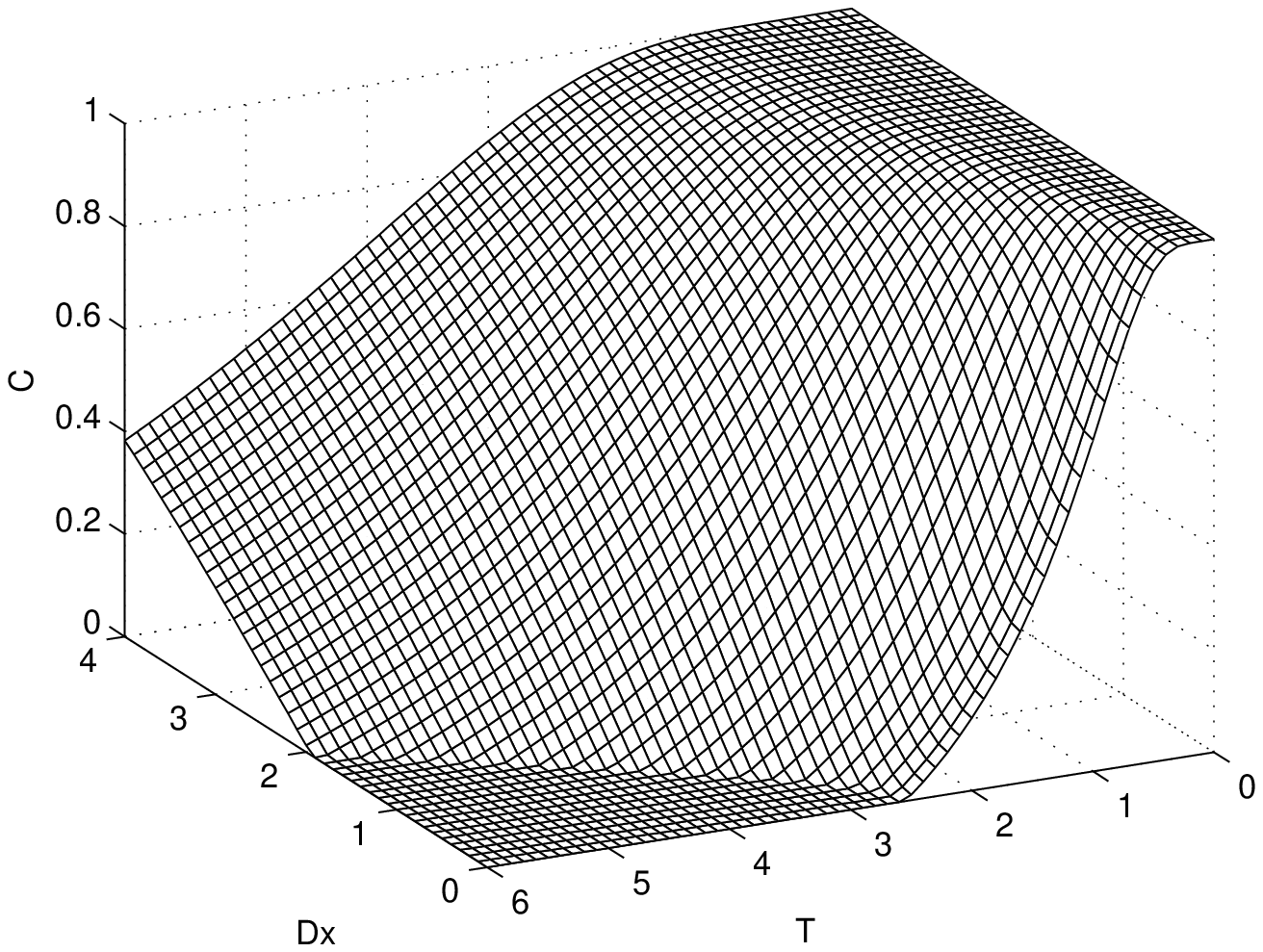}
\caption{The concurrence is plotted versus $T$ and $D_{x}$ where
the coupling constants $J=1$ and $J_{z}=0.2$.} \label{fig4}
\end{figure}

\begin{figure}[tbp]
\includegraphics[scale=0.5,angle=0]{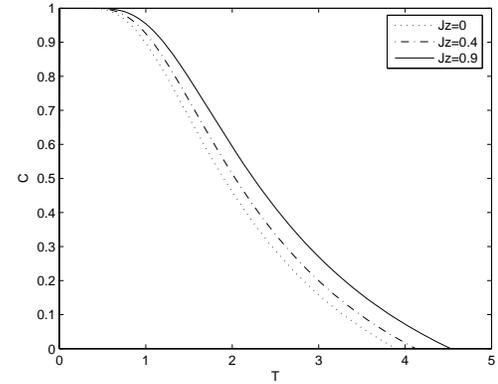}
\caption{The concurrence versus $T$ for different $J_{z}$ in the
system with x-component parameter $D_{x}$, here $J=1$ and
$D_{x}=1$.} \label{fig5}
\end{figure}

To analyze the role of parameters and the variation of the
entanglement, we restrict $D_{x}>0$, $J_{z}>0$ and $J>0$. In Fig.
4, the concurrence is plotted versus $T$ and $D_{x}$ when the
coupling constants $J=1$ and $J_{z}=0.2$. It is evident that
increasing temperature will decrease the entanglement, and
increasing $D_{x}$ will enhance the entanglement and increase the
critical temperature $T_{c}^{'}$ which is determined by Eq. (16).
Fig. 5 demonstrates that the concurrence versus $T$ for different
$J_{z}$ with x-component parameter $D_{x}=1$ and $J=1$, it is easy
to find that increasing $J_{z}$ can increase the critical
temperature and enhance the entanglement for a certain
temperature. So $D_{x}$ and $J_{z}$ are both efficient control
parameters of entanglement, too.

\section{The comparison between the two DM
interaction component parameters}From Section II and Section III,
we know that the x-component parameter $D_{x}$ and the z-component
parameter $D_{z}$ of DM interaction have the similar qualities.
They are both efficient control parameters of entanglement,
increasing them can enhance the entanglement or increase the
critical temperature to slow down the decrease of entanglement.

\begin{figure}[tbp]
\includegraphics[scale=0.5,angle=0]{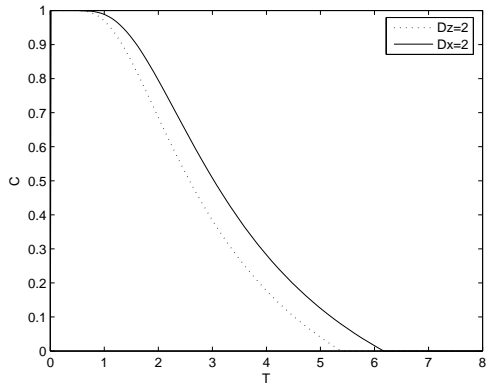}
\caption{The concurrence is plotted as a function of the
temperature $T$ for $D_{z}=2$ and $D_{x}=2$, here $J=1$ and
$J_{z}=0.2$.} \label{fig6}
\end{figure}

In this section, we mainly analyze the differences between
x-component parameter and z-component parameter. We note that
there is less disentanglement region in Fig. 4 than Fig. 1, where
$J=1$ and $J_{z}=0.2$, and the temperature and the spin-orbit
coupling parameter have the same range. We also see that
increasing x-component parameter $D_{x}$ can make the entanglement
increase more rapidly, for example, when $T=6$ the concurrence
increases more rapidly in Fig. 4 than Fig. 1. These phenomena
denote that x-component parameter $D_{x}$ has a more remarkable
influence than the z-component parameter $D_{z}$. In Fig. 6, the
concurrence is plotted as a function of the temperature $T$ for
$D_{z}=2$ and $D_{x}=2$ with $J=1$ and $J_{z}=0.2$. It is easy to
see that for the same $D_{x}$ and $D_{z}$, x-component parameter
$D_{x}$ has a higher critical temperature and more entanglement
for a certain temperature than z-component parameter $D_{z}$. We
show distinctly the differences between different DM interaction
parameters in Fig. 6. So using the DM interaction parameters in
different directions, we can increase the entanglement and the
critical temperature with different efficiencies.

\section{Discussion}
The entanglement of a two-qubit Heisenberg XXZ system with
different DM interaction parameters have been investigated. The DM
interaction parameter and coupling coefficient $J_{z}$ are
efficient control parameters of the entanglement. By increasing
the parameters, we can enhance the entanglement or increase the
critical temperature to slow down the decreace of the
entanglement. In addition, we have also investigated the
differences between x-component DM interaction parameter $D_{x}$
and z-component parameter $D_{z}$. Entanglement can be increased
more rapidly by increasing $D_{x}$ than $D_{z}$. When $D_{x}$ and
$D_{z}$ have the same value, $D_{x}$ has a higher critical
temperature than $D_{z}$. Thus, by changing the direction of DM
interaction, we can obtain a more efficient control parameter to
increase the entanglement and the critical temperature.

\begin{acknowledgments}
This work is supported by National Natural Science Foundation of
China (NSFC) under Grant Nos: 60678022 and 10704001, the
Specialized Research Fund for the Doctoral Program of Higher
Education under Grant No. 20060357008, Anhui Provincial Natural
Science Foundation under Grant No. 070412060, the Talent
Foundation of Anhui University, and Anhui Key Laboratory of
Information Materials and Devices (Anhui University).
\end{acknowledgments}


\begin{thebibliography}{99}
\bibitem{1} C. H. Bennett and S. J. Wiesner, Phys. Rev. Lett. 69 (1992) 2881; M. Murao, D. Jonathan, M. B. Plenio and V. Vedral, Phys. Rev. A 59 (1999) 156.
\bibitem{2} M. A. Neilsen and I. L. Chuang, Quantum computation and quantum information (Cambridge University Press), Cambridge, UK, 2000.
\bibitem{3} X. Wang, Phys. Rev. A 64 (2001) 012313.
\bibitem{4} K. M. OConnor and W. K. Wootters, Phys. Rev. A 63 (2001) 052302.
\bibitem{5} G. L. Kamta and A. F. Starace, Phys. Rev. Lett. 88 (2002) 107901.
\bibitem{6} D. V. Khveshchenko, Phys. Rev. B 68 (2003) 193307.
\bibitem{9} T. Senthil, J. B. Marston and Matthew P. Fisher, Phys. Rev. B 60 (1999) 4245.
\bibitem{10} M. Nishiyama, Y. Inada and Guo-qing Zheng, Phys. Rev. Lett. 98 (2007) 047002.
\bibitem{11} B. Trauzettel, Denis V. Bulaev, Daniel Loss and Guido Burkard, Nature Phys. 3 (2007) 192.
\bibitem{12} R. Hanson, L. P. Kouwenhoven, J. R. Petta, S. Tarucha and L. M. K. Vandersypen, Rev. Mod. Phys. 79 (2007) 1217.
\bibitem{13} F. Bodoky and M. Blaauboer, Phys. Rev. A 76 (2007) 052309.
\bibitem{14} D. Porras and J. I. Cirac, Phys. Rev. Lett. 92 (2004) 207901.
\bibitem{15} X. X. Yi, H. T. Cui and L. C. Wang, Phys. Rev. A 74 (2006) 054102.
\bibitem{7} Y. Sun, Y. Chen and H. Chen, Phys. Rev. A 68 (2003) 044301.
\bibitem{8} G. F. Zhang and S. S. Li, Phys. Rev. A 72 (2005) 034302.
\bibitem{27} M. Asoudeh and V. Karimipour, Phys. Rev. A 71 (2005) 022308.
\bibitem{28} X. Q. Su and A. M. Wang, Phys. Lett. A 369 (2007) 196.
\bibitem{29} Z. N. Gurkan and O. K. Pashaev, e-print quant-ph/0705.0679 and quant-ph/0804.0710.
\bibitem{20} Y. Yeo, Phys. Rev. A 66 062312 (2002).
\bibitem{22} X. Hao and S. Q. Zhu, Phys. Lett. A 338 (2005) 175.
\bibitem{23} L. Campos Venuti, S. M. Giampaolo, F. Illuminati, and P. Zanardi, Phys. Rev. A 76 (2007) 052328.
\bibitem{21} G. F. Zhang, Phys. Rev. A 75 (2007) 034304.
\bibitem{30} F. Kheirandish \emph{et al.}, Phys. Rev. A 77 (2008) 042309.
\bibitem{16} D. Loss and D. P. DiVincenzo, Phys. Rev. A 57 (1998) 120.
\bibitem{17} D. A. Lidar, D. Bacon, and K. B. Whaley, Phys. Rev. Lett. 82 (1999) 4556.
\bibitem{18} D. P. Divincenzo, D. Bacon, J. Kempe, G. Burkard, and K. B. Whaley, Nature (London) 408 (2000) 339.
\bibitem{19} L. F. Santos, Phys. Rev. A 67 (2003) 062306.
\bibitem{24} M. A. Nielsen, arXiv:quant-ph/0011036.
\bibitem{25} S. Hill and W. K. Wootters, Phys. Rev. Lett. 78 (1997) 5022.
\bibitem{26} W. K. Wootters, Phys. Rev. Lett. 80 (1998) 2245.
\end{thebibliography}
\end{document}